\documentclass[a4paper]{jpconf}
\usepackage{graphicx}
\usepackage{amsfonts}
\usepackage{amsmath}
\usepackage{amssymb}
\usepackage[english]{babel}
\usepackage[latin1]{inputenc}
\begin{document}
\title{A non-relativistic magnetized vector boson gas at any temperature}

\author{LC. Su\'arez-Gonz\'alez$^1$, G. Quintero Angulo$^2$,
    A. P\'erez Mart\'{\i}nez$^{1}$ and H. P\'erez Rojas$^1$}

\address{$^1$ Instituto de Cibern\'etica Matem\'atica y F\'{\i}sica (ICIMAF),
    Calle E esq 15 No. 309 Vedado, La Habana, CP 10400, Cuba\\
$^2$ Facultad de F\'isica, Universidad de La Habana,San L\'azaro y L, CP 10400, Cuba}

\ead{lismary@icimaf.cu}

\begin{abstract}
We study the thermodynamic properties of a neutral vector boson gas in presence of a constant magnetic field, by means of a semi-classical approach that allows to introduce the spin in the non-relativistic spectrum of the bosons.  Bose-Einstein condensation is obtained and it turns out to depend on all the parameters involved in the problem: temperature, particle density and magnetic field. A spontaneous magnetization appears at low temperature as a consequence of the condensed state. The axial symmetry imposed in the system by the magnetic field presence, splits the pressure in two components, one along and another perpendicular to the magnetic axis. Under certain conditions, the perpendicular pressure becomes negative signaling that the system undergoes a transversal magnetic collapse. The spontaneous magnetization might be useful to model magnetic field production inside compact stars, while the negative pressures imposes certain limits to the temperatures and densities needed inside these objects to support a given magnetic field.
\end{abstract}

\section{Introduction}

Neutron stars (NS) are the smaller and denser objects known in our Universe so far \cite{Camenzind}. Although they have been widely studied, there are no consensus about their internal composition, mainly because matter at such extreme conditions cannot been obtained yet at lab. Nevertheless, a lot of exotic particles and phases have been conjecture to exist in NS interiors \cite{Page:2011yz}. In particular, it has been proposed that, at some stage of the NS evolution, it might contain certain amount of bosons formed up by the pairing of neutrons and protons in the crust and core, and electron and positrons in the magnetosphere \cite{Page:2011yz,Shternin:2010qi,Chavanis:2011}. Given the high magnetic fields present in most NS, the magnetic properties of these particles are expected to be relevant for the structure, composition and other physical phenomena related to these compact objects.

Our main purpose is to study the thermodynamic properties of a magnetized gas of neutral spin-1 particles (neutron-neutron or electron-positron spin parallel paired), with the aim of providing equations of state that allows more accurate descriptions of magnetic neutron stars. However, note that the occurrence of Bose-Einstein condensation, one of the most outstanding properties of bosonic systems, in presence of magnetic fields is of interest in others areas, such as condensed matter and particle physics \cite{berman2010bose,2005PhRvL95e0401H}, in which this work might be also applied.

The thermodynamics of relativistic magnetized boson gases has been tackled in \cite{ROJAS1996148, Khalilov1997,Khalilov1999,PEREZROJAS2000,angulo2017thermodynamic} for the low temperature regime ($T<<m$). In this sense, our work can be seen as an extension of the former papers, because
Bose-Einstein condensation (BEC), magnetization and the anisotropic equations of state (EoS) are obtained at any temperature, although so far we are taken the bosons as non relativistic. The extension of the study at any temperature is convenient mainly for two reasons. First, it provides an easy way to check the computations consistency since in the high temperature region the well known classical behaviour has to be recovered. And second, this extension eliminates the restriction that $T$ has to be much less than the particle mass, that depending on the kind of bosons might be in contradiction with the temperatures expected inside the NS.

On the other hand, a non-relativistic magnetized spin-1 boson gas has been investigated in \cite{yamada1982thermal}. In what concerns BEC and magnetic properties, our results are in general agreement with those shown in the above mentioned paper. However, in \cite{yamada1982thermal} the breaking of the SO$3$ rotational symmetry produced by the magnetic field is never taken into account, in spite of this is a very relevant feature that splits the pressure in two components, perpendicular and parallel to the magnetic field direction, and may cause instabilities in the system \cite{Chaichian:1999gd}. In order to provide a complete description of the non-relativistic magnetized spin-1 gas, we devote a section to the analysis of the stability and the anisotropic pressures.

The paper is organized as follows. In Section \ref{sec1}, the thermodynamical potential of the non-relativistic spin-1 gas under the action of an external magnetic field is computed. Section \ref{sec2} is devoted to Bose-Einstein condensation and magnetic properties. Anisotropic EoS and magnetic instabilities are discussed in Section \ref{sec3}. Finally, in Section \ref{sec4} concluding remarks are given.

\section{Thermodynamical potential of a non-relativistic vector boson gas interacting with a magnetic field}	
\label{sec1}

In this section we compute the thermodynamical potential of an ideal non-relativistic neutral vector boson gas interacting with a constant and uniform magnetic field $\overrightarrow{B}=(0,0,B)$. The spectrum of the bosons is $\varepsilon(p,S_z)=\vec{p}^{\:2}/2m -S_{z}\kappa B$, being $m$ the mass of the particles, $\kappa$ their magnetic moment, $\vec{p}$ the momentum and $S_ {z} = -1,0,+1$ the projection of the spin in the z direction. For such a gas, the density of states $g(\epsilon)$ reads

\begin{eqnarray}\label{eq1}
g(\epsilon)&=&\frac{4\pi V}{(2\pi \hbar)^3}
\sum_{S_z=-1,0,1}  \int_0^\infty dp\;p^2 \delta\left(\epsilon-\varepsilon(p,S_z)\right)\nonumber \\
&=&\frac{4\pi V}{(2\pi \hbar)^3}
\sum_{S_z=-1,0,1}  \int_0^\infty dp\;p^2 \delta\left(\epsilon-\frac{\vec{p}^{\:2}}{2m}+ S_z\kappa B\right)\nonumber \\
&=& \frac{4\pi V}{(2\pi \hbar)^3} \bigg[ \int_0^\infty dp\;p^2 \delta\left(\epsilon-\frac{\vec{p}^{\:2}}{2m}-\kappa B\right)
+ \int_0^\infty dp\;p^2 \delta\left(\epsilon-\frac{\vec{p}^{\:2}}{2m}\right)\\
&+&    \int_0^\infty dp\;p^2 \delta\left(\epsilon-\frac{\vec{p}^{2}}{2m}+\kappa B\right)\bigg]. \nonumber
\end{eqnarray}

\noindent where $\epsilon$ is the energy. Note that Eq.~(\ref{eq1}) can be separated in three terms, each one corresponding to a specific spin state. After doing the integration $g(\epsilon)$ becomes
\begin{eqnarray}\label{eq3}
g(\epsilon)=  \frac{4\pi mV}{(2\pi \hbar)^3} \left[ \sqrt{2m(\epsilon-\kappa B)}+ \sqrt{2m \epsilon} + \sqrt{2m(\epsilon +\kappa B)} \right]. \\ \nonumber
\qquad \qquad \qquad \qquad
\end{eqnarray}

\noindent Using  Eq.~(\ref{eq3}) the thermodynamical potential $\Omega(\mu,T,B)$ might be written as

\begin{equation}\label{eq4}
\Omega(\mu,T,B)=\frac{T}{V}\int_0^\infty d\epsilon g(\epsilon)\;ln (\emph{f}_{BE} (\epsilon,\mu)), \;\;\forall\;\mu<\epsilon,
\end{equation}

\noindent with $\emph{f}_{BE}(\epsilon,\mu)=\left[1-e^{\beta(\mu-\epsilon)}\right]^{-1}$ being the Bose-Einstein distribution function, $\mu$ the chemical potential, $T$ the absolute temperature and $\beta=1/T$. As the density of states, $\Omega(\mu,T,B)$ can be divided in three terms $\Omega(\mu,T,B)=\Omega_{-}(\mu,T,B)+\Omega_{0}(\mu,T,B)+\Omega_{+}(\mu,T,B)$, where $\Omega_{-}(\mu,T,B)$, $\Omega_{0}(\mu,T,B)$ and $\Omega_{+}(\mu,T,B)$ corresponds to the states with $S_z=-1$, $S_z=0$ and $S_z=1$ respectively. Integrating over the energy in Eq.~(\ref{eq4}) one gets

\begin{eqnarray}\label{eq12}
\Omega_{-}(\mu,T,B)&=&- \frac{T}{\lambda^3} g_{5/2}(z_{-}),\\
\Omega_{0}(\mu,T,B)&=&- \frac{T}{\lambda^3} g_{5/2}(z),\label{eq1a2} \\
\Omega_{+}(\mu,T,B)&=& - \frac{T}{\lambda^3} g_{5/2}(z_{+}),\label{eq12b}
\end{eqnarray}

\noindent where $g_{5/2}(x) = \sum_{l=1}^\infty x^l/l^{5/2}$ is the polylogarithmic function of order $5/2$, $\lambda=\sqrt{2\pi/mT}$ is the thermal wavelength, $z=e^{\mu/T}$ is the fugacity and $z_{\sigma}=z e^{\sigma \frac{\kappa B}{T}}$ where $\sigma=-,+$. With the use of Eqs.~(\ref{eq12})-(\ref{eq12b}), all the thermodynamic magnitudes -particle density, magnetization, pressures, etc.-, and the contributions to them of each spin state, can be determined.

\section{Bose-Einstein condensation and magnetic properties}
\label{sec2}

We start the study of  Bose-Einstein condensation (BEC) by  calculating the particle number density $\rho$

\begin{equation}\label{eq17}
\rho = \rho_{gs}(T,B) -\frac{\partial \Omega(\mu, T,B)}{\partial \mu},
\end{equation}

\noindent where $\rho_{gs}$ stands for the particles in the condensate, i.e. in the ground state, and is such that $\rho_{gs}(T,B)=0$ if $T \geq T_c$ and $\rho_{gs}(T,B)>0$ if $T<T_c$, being $T_c$ the BEC critical temperature. After doing the derivative in Eq.~(\ref{eq17}), the particle density takes the form

\begin{eqnarray}\label{eq18}
\rho &=&\rho_{gs}(T,B) + \rho_{-}(\mu,T,B)+\rho_{0}(\mu,T,B)+\rho_{+}(\mu,T,B),\\
\rho &=& \rho_{gs}(T,B) + \frac{g_{3/2}(z_{-}) }{\lambda^3}+ \frac{g_{3/2}(z)}{\lambda^3}+ \frac{g_{3/2}(z_{+})}{\lambda^3}.\label{eq18a}
\end{eqnarray}

\begin{figure}[h!]
\centering
		\includegraphics[width=0.49\linewidth]{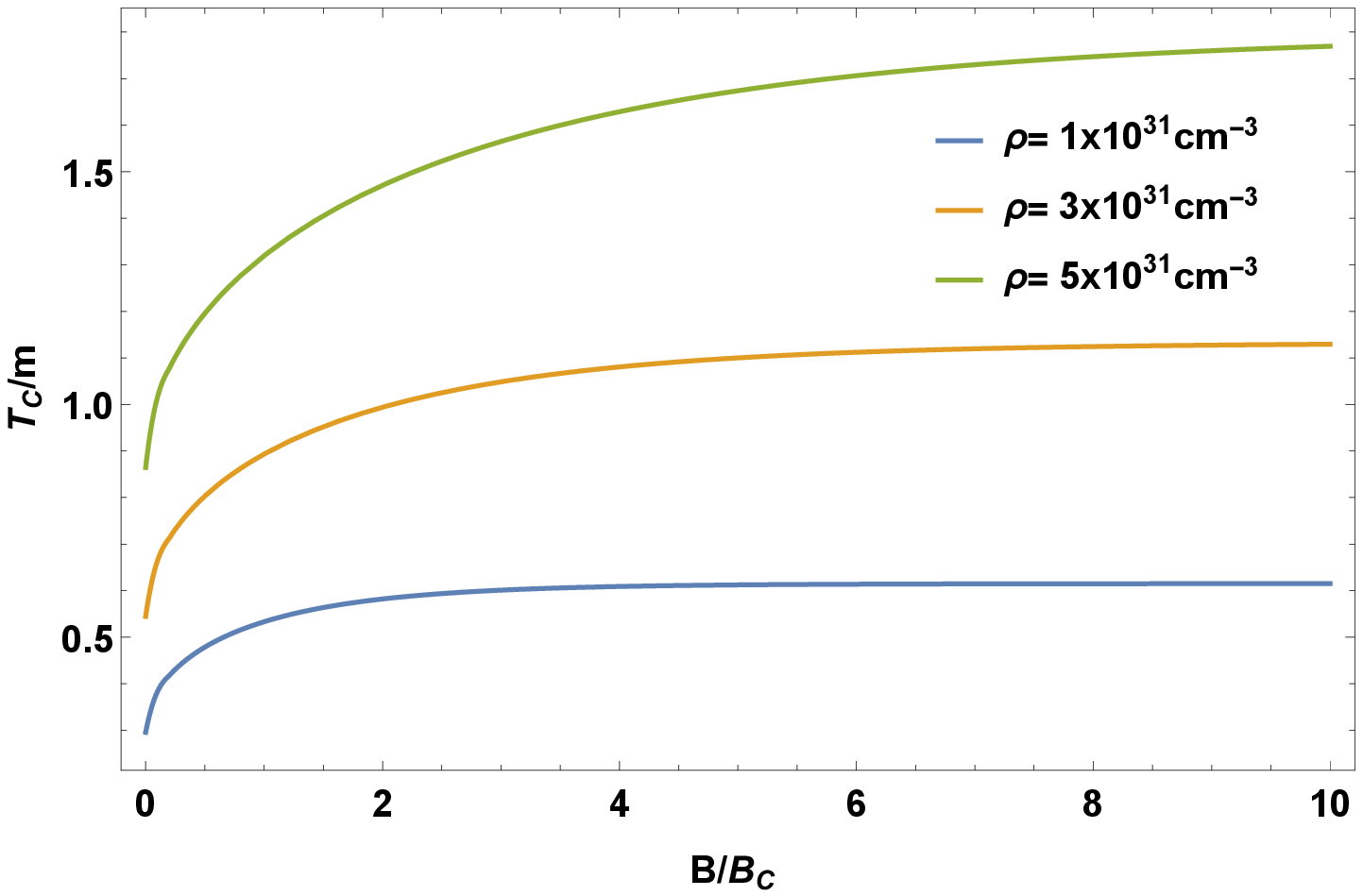}
        \includegraphics[width=0.49\linewidth]{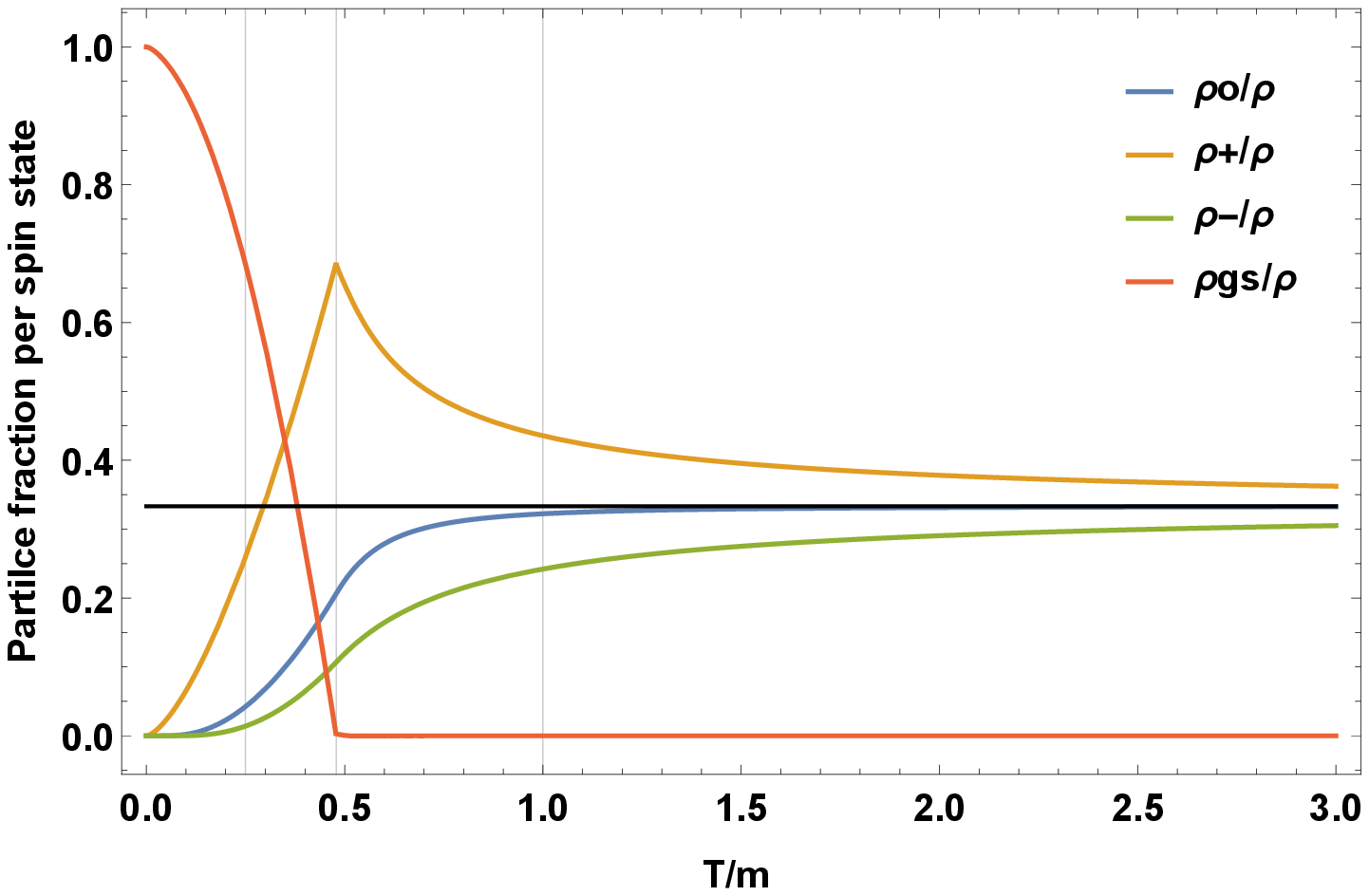}
		\caption{\label{f1}Left panel, BEC critical temperature as a function of the magnetic field for several values of the particle density. Right panel, particle density per spin state as a function of temperature for $\rho = 10^{31}$cm$^{-3}$ and $B = 0.5 B_c$.}
	\end{figure}

Using  Eq.~(\ref{eq18a}) we compute $T_c$ numerically as a function of $\rho$ and $B$. To do so, it is enough to recall that when the transition to the BEC begins, $\rho_{gs} = 0$ and the chemical potential $\mu$ equals the particle's rest energy  $\varepsilon_{min} = -\kappa B$. The result is depicted in left panel of Fig.\ref{f1}, where $T_c$ is plotted as a function of the magnetic field for several fixed values of the particle density. The results are given for bosons of mass $m=2m_e$ and $\kappa = 2 \mu_B$, where $m_e$ is the electron mass and $\mu_B$ the Bohr magneton. $B_c=m/2\kappa$ is the magnetic field value at which the magnetic energy becomes comparable to the mass of the particles.

The left panel of Fig.\ref{f1} shows how the BEC critical temperature grows starting from its $B=0$ value

\begin{equation}\label{eq21a}
T_{c}(0)= \frac{2\pi}{m} \left( \frac{\rho}{3 g_{3/2}(1)}  \right)^{2/3},
\end{equation}
\noindent until saturation when $B \rightarrow \infty$ for wich
\begin{eqnarray}\label{eq21b}
T_{c}(\infty)=\frac{2\pi}{m} \left( \frac{\rho}{ g_{3/2}(1)}  \right)^{2/3}.
\end{eqnarray}
These extreme values of the BEC critical temperature are in agreement with the ones obtained in \cite{yamada1982thermal}. On the other hand, the asymptotic behavior of $T_c$ with growing $B$ constitutes a major difference between this case and the relativistic one, in which the BEC critical temperature always increases with $B$ and diverges when $B \rightarrow B_c$ \cite{angulo2017thermodynamic}.

The saturation of $T_c$ in the strong magnetic field region give us valuable information about the influence of $B$ on the BEC of non-relativistic particles. For magnetic field values before saturation, increasing $B$ increases $T_c$ in a noticeable way, driving the system to condensation. But when the magnetic field reaches the saturated region, further changes barely affects $T_c$, in spite of what we should not forget that $T_{c}(\infty)$ is bigger than $T_{c}(0)$. Therefore, one can conclude that in general, for a fixed particle density, the magnetic field effect on the BEC is to increase the critical temperature, and the weaker the field, the more sensitive is the system to change on its values. Note also that, although the BEC critical temperature increases with the particle density, the ratio between the two extreme values $T_c(\infty)/T_c(0) = \sqrt[3]{9} $ is constant.

Furthermore, it is of interest to look at the behavior of the particle fraction in the BEC $\rho_{gs}/\rho$ and per spin state $\rho_{\sigma}/\rho$, ${\sigma = -,0,+}$ as a function of temperature. This is shown in right panel of Fig.~\ref{f1}. In the high temperature region, $T \gg m$, $\rho_{gs}=0$ and $\rho_{\sigma}/\rho \rightarrow 1/3$ for all $\sigma$, because this is a temperature dominating region in which thermal disorder rules. When $T$ decreases, the magnetic field ordering effect begins to be noticed and the fraction of particles with spin aligned to the field $\rho_{+}/\rho$ becomes the higher one. This behavior continues through the low temperature region $T \ll m$ being the next appreciable change when $T = T_c$. At this point the fraction of particles in the BEC becomes non zero and increases with decreasing temperature until it equals $1$ at $T=0$, where $\rho_{\sigma}/\rho=0$ for all $\sigma$. Nevertheless, since the condensate in the ground state is determined by $S_z = 1$, it is expected that, for a magnetic boson gas, a non zero magnetization exists even in the absence of magnetic field. To check on this, we computed the gas magnetization

\begin{figure}[h!]
\centering
		\includegraphics[width=0.49\linewidth]{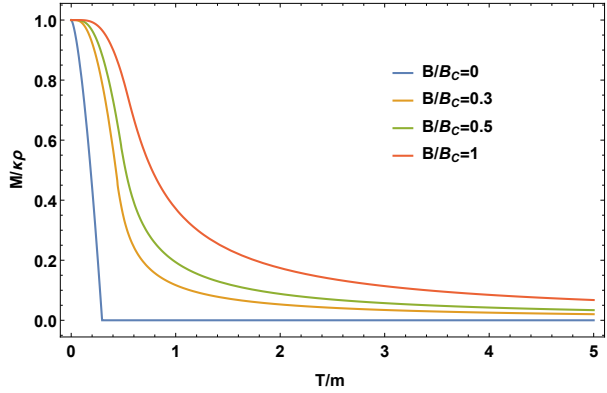}
     	\caption{\label{f3}Magnetization as a function of temperature for $\rho=10^{31}$cm$^{-3}$ and several values of the magnetic field.}
	\end{figure}

\begin{equation}\label{eq28}
M= \kappa\rho_{gs} -\left(\frac{\partial \Omega}{\partial B}\right) =  \kappa(\rho_{gs} + \rho_{+})-\kappa \rho_{-},
\end{equation}

\noindent and plotted it in Fig.~\ref{f3} for $\rho=10^{31}$cm$^{-3}$ and several values of $B$, including $B=0$. The curves on this graph are in accordance with the ones in Fig.~\ref{f1} (right panel): $M \rightarrow 0$ for high $T$, while $M \rightarrow \kappa \rho$ when $T \rightarrow 0$. The outstanding feature of Fig.~\ref{f3} is that this happens also for $B=0$, i.e. the gas shows a spontaneous magnetization that in this case is not due to an interaction between the spin of the particles but to the BEC. Note that, for $B=0$, $M \neq 0$ only when $T < T_c$. This can be also obtained by direct substitution of $B=0$ in Eq.~(\ref{eq28}), which gives $M(B=0)=\kappa \rho_{gs}$. This result acquires astrophysical relevance since the spontaneous magnetization might provide a magnetic field source for the interior of compact objects.

\section{Equations of state}
\label{sec3}

To compute the EoS, the breaking of the SO3 rotational symmetry produced by the magnetic field in the system has to be taken into account  \cite{Chaichian:1999gd}. As a consequence of that, the energy-momentum tensor of the particles becomes anisotropic and the pressure splits in two components, one parallel, $P_{\parallel}$, and the other perpendicular, $P_{\perp}$, to the magnetic axis. Therefore, for a magnetized quantum gas of either neutral or charged, bosonic or fermionic, particles the EoS read

\begin{eqnarray}\label{EoS}
	E &=&\Omega + \mu \rho -T S, \label{energia}\\
	P_{\parallel} &=&  -\Omega, \label{presionpara} \\
	P_{\perp}&=& -\Omega -{\mathcal M} B, \label{presionper}
\end{eqnarray}

\noindent where $S= -\left(\partial \Omega_{B}/\partial T\right)_{T,V}$ is the entropy of the system, that in the present case is

\begin{eqnarray}\label{eq24}
S= -\frac{5}{2}\frac{\Omega}{T} - \mu \frac{\rho-\rho_{gs}}{T} - \kappa B \frac{\rho_+ - \rho_-}{T}.
\end{eqnarray}
Combining Eq.~(\ref{eq24}) with Eqs.~(\ref{EoS}) and (\ref{eq12}), we get for the energy density

\begin{eqnarray}\label{eq25}
E=\frac{3}{2} P_{\parallel} - \kappa B (\rho_{gs} + \rho_{+} - \rho_{-}).
\end{eqnarray}

\begin{figure}[h]
\centering
		\includegraphics[width=0.49\linewidth]{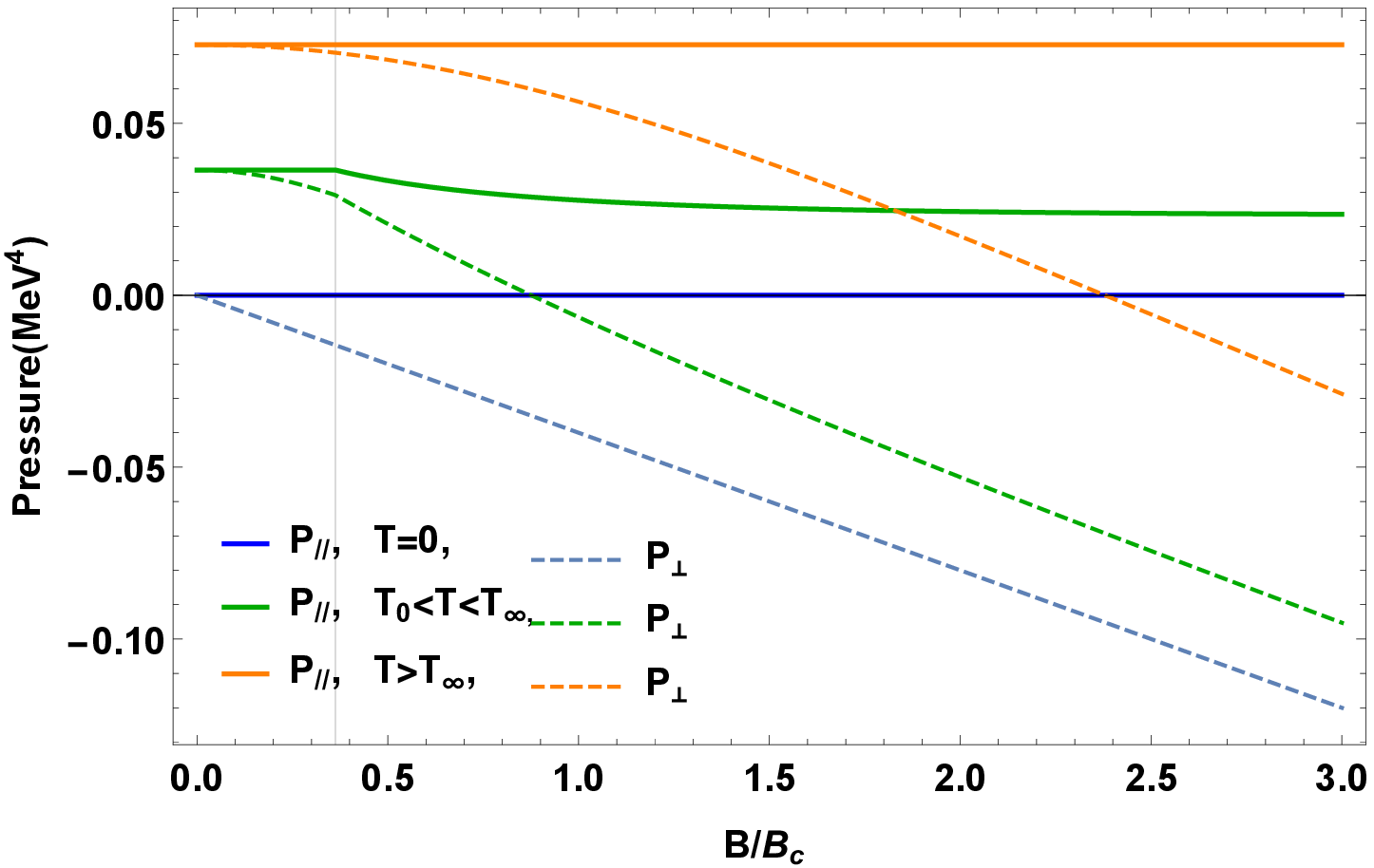}
        \includegraphics[width=0.49\linewidth]{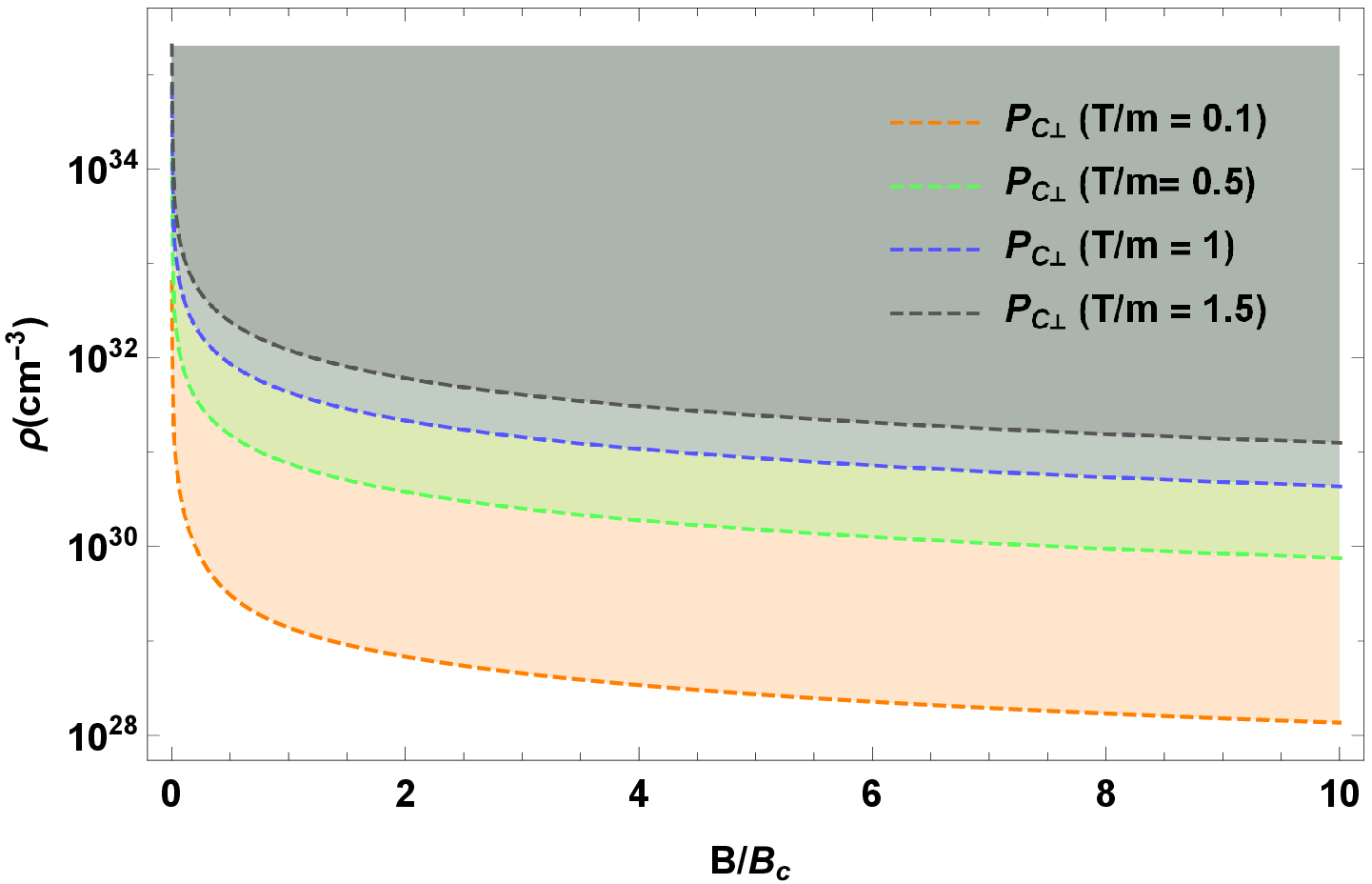}
		\caption{\label{f5} Left plot, the pressures as a function of the magnetic field for several values of the temperature and $\rho = 10^{31}$cm$^{-3}$. Right plot, phase diagram for the transverse magnetic collapse in the particle density-magnetic field plane for several values of the temperature; the dashed lines stands for the solution of $P_{\perp}(T,B,\rho)=0$ at a fixed temperature; for each temperature, the instability region has been shadowed.}
\end{figure}

\noindent While from Eqs.~(\ref{presionpara}) and (\ref{presionper}) one can see that the pressures are expressed in terms of quantities $\Omega$ and $M$ computed above. The EoS, Eqs.~(\ref{presionpara}), (\ref{presionper}) and (\ref{eq25}), can be used to model magnetized compact objects composed partially of entirely of magnetized bosons. However, care must be taken, since, as shows Fig.~\ref{f5}, in dependence of the temperature, the magnetic field and the particle density, the perpendicular pressure might be negative.

Left panel of Fig.~\ref{f5} shows the pressures as a function of the magnetic field for $\rho=10^{31}$cm$^{-3}$ and several values of the temperature. At zero magnetic field, $P_{\parallel} = P_{\perp}$ and the system is isotropic. If $B \ne 0$ the difference between the pressures increases when decreasing the temperature or augmenting the magnetic field. In fact, it is evident from the plot that the magnetic field barely affects the parallel pressure, while its contribution is very important to the perpendicular one. From a microscopic point of view, this is because the magnetic field diminishes the perpendicular momentum of the particles, but does not affect the parallel one. Macroscopically, this effect is expressed in the subtractive term $- M B$ (note that $M>0$) that appears in the perpendicular pressure. That is why $P_{\perp}$ is always less than $P_{\parallel}$ and decreases until becoming zero or even negative with the increase of the magnetic field. Since the effect of a negative perpendicular pressure is to push the particles inwards to the magnetic axis, this can be interpreted as the system becoming unstable. This kind of instability has been previously observed in other magnetized quantum gases and it is known as transverse magnetic collapse \cite{Chaichian:1999gd,Aurora2003EPJC,Felipe:2002wt,Elizabeth,Quintero2017AN}.

In right panel of Fig.~\ref{f5} a phase diagram for the collapsed/non collapsed gas is drawn in the particle density-magnetic field plane, for several values of the temperature. The dashed lines stand for the solution of $P_{\perp}(T,B,\rho)=0$ at a fixed temperature. The gas is stable in the region under the lines and unstable over them. At $T \neq 0$ and $B = 0$, an infinite number of particles is needed for the system to become unstable, an expected result since the magnetic field is the cause of the instability. When $T=0$ and $B \neq 0$, $P_{\parallel} = 0$ and $P_{\perp} = -M B$ is always negative, i.e. the pure condensed state is unstable for any non-zero magnetic field value. In the case that both the temperature and the magnetic field are different from zero decreasing the temperature favors the collapse, as well as increasing the magnetic field. In addition, Fig.~\ref{f5} also shows that the collapse is favored by particle density augmentation, because the denser the gas, the higher are $M$ and $T_c$. In consequence, for a fixed temperature and magnetic field increasing $\rho$ decreases the thermal pressure ($-\Omega(\mu,T,B)$) and increases the magnetic pressure ($-MB$) driving $P_{\perp}$ to negative values. In this sense, the collapse imposes an upper limit to the particle densities that can exist inside compact stars with a given magnetic field. Nevertheless, we would like to remark that since a compact star has an heterogeneous composition, the occurrence or not of the collapse will depend on the pressures, and therefore, on the magnetic response, of all the present species.

\section{Conclusions} \label{sec4}

We have studied the properties of the BEC, magnetization and EoS of a magnetized vector boson gas at any temperature. The bosons are considered non-relativistic and the spin is included under a semi-classical approach.

The phase transition to the BEC depends on the temperature, the particle density and the magnetic field, in a way that increasing the particle density or the magnetic field, as well as decreasing the temperature, drives the system to the condensed state. For a fixed particle density, the critical temperature grows with the magnetic field from its value at $B=0$, $T_c(0)$ to the limiting value $T_c(\infty)$, being the ratio between this extreme values independent of density.

Below the BEC critical temperature, the gas shows a spontaneous magnetization. This magnetization is not due to a spin coupling between the particles, but to the fact that bosons in the condensed phase are in the state of lowest energy. For a system of magnetic bosons this state is such as all the particles have the same spin projection. The ability of this system to be spontaneously magnetized might be connected to magnetic field production in astrophysical objects.

As happens for other magnetized quantum gases, the magnetic field presence imposes its axial symmetry to the system and separates the pressures in two components, one along and the other perpendicular to the magnetic axis. Under certain conditions, the perpendicular pressure might vanish or be negative and the system becomes unstable. As was shown, this instability is caused by the magnetic field, while the temperature opposes it. In consequence, the cooler the matter, the more susceptible it is to undergo a transverse magnetic collapse. On the other hand, increasing the particle density also destabilizes the gas. This imposes an upper limit to the boson densities allowed in astronomical objects for a given temperature and magnetic field.

\section*{Acknowledgments}
This work have been supported by the grant No. 500.03401 of PNCB-MES, Cuba. A.P.M acknowledges the support of the  organizing committee of IARD 2018 who allowed her participate in the nice IARD conference. She  also thanks  the Caribbean network of OEA-ICTP for supporting her visit to CINVESTAV, Mexico, where this work was finished. G.Q.A. express gratitude to the Service de Coopération et d'Action Culturelle (SCAC) of the Embassy of France in Cuba and to the Federated Institutes Programme of the ICTP, Italy, for support during the final stage of this work. L.C.S.G. and G.Q.A. acknowledges D. Alvear Terrero for her help in some calculations and in the elaboration of the figures. The authors would also like to express their gratitude to the referees for their comments and useful suggestions.

\section*{References}
\bibliographystyle{iopart-num}
\bibliography{bibL}
\end{document}